\documentclass[aps,prd,twocolumn,tightenlines,superscriptaddress,
nofootinbib,preprintnumbers]{revtex4-1}

\usepackage{amsmath}
\usepackage{amssymb}
\usepackage{latexsym}
\usepackage{graphicx}
\usepackage{slashed}
\usepackage{pifont}
\usepackage{hyperref}
\usepackage{color}
\usepackage{cleveref}
\usepackage{multirow}
\usepackage[usenames,dvipsnames]{xcolor}

\hypersetup{colorlinks=true,citecolor=[rgb]{0,0,0.8},
linkcolor=[rgb]{0,0,0.8},urlcolor=[rgb]{0,0,0.8}}

\newcommand{\Blue}[1]{\textcolor{black}{#1}}
\definecolor{joelred}{rgb}{0.5,0,0}

\begin{document}
\preprint{\hspace{13.45cm}IQuS@UW-21-105} 

\title{Parton Distributions on a Quantum Computer}
\author{Jiunn-Wei Chen}
\email{jwc@phys.ntu.edu.tw}
\affiliation{Department of Physics and Center for Theoretical Physics,
National Taiwan University, Taipei, Taiwan 106}
\affiliation{Physics Division, National Center for Theoretical Sciences, Taipei 10617, Taiwan}
\affiliation{InQubator for Quantum Simulation (IQuS), Department of Physics, University of Washington, Seattle, WA 98195, USA.}
\author{Yu-Ting Chen}
\email{asdasdasdasdasd2578@gmail.com}
\author{Ghanashyam Meher}
\email{ghanashyam@phys.ntu.edu.tw}
\affiliation{Department of Physics and Center for Theoretical Physics,
National Taiwan University, Taipei, Taiwan 106}
\affiliation{Physics Division, National Center for Theoretical Sciences, Taipei 10617, Taiwan}


\begin{abstract} 
We perform the first quantum computation of parton distribution function (PDF) with a real quantum device by calculating the PDF of the lightest positronium in the Schwinger model with IBM quantum computers. The calculation uses 10 qubits for staggered fermions at five spatial sites and one ancillary qubit. The most critical and challenging step is to reduce the number of two-qubit gate depths to around 500 so that sensible results start to emerge. 
The resulting lightcone correlators have excellent agreement with the classical simulator result in central values, although the error is still large.
Compared with classical approaches, quantum computation has the advantage of not being limited in the accessible range of parton momentum fraction $x$ due to renormalon ambiguity, and the difficulty of accessing non-valence partons. A PDF calculation with 3+1 dimensional QCD near $x=0$ or $x=1$ will be a clear demonstration of the quantum advantage on a problem with great scientific impact. 

\end{abstract}

\maketitle
\section{Introduction}

Parton distribution functions (PDFs) describe how momentum and spin of a hadron are distributed among the partons (i.e., quarks and gluons) in the infinite-momentum frame of the hadron.
They are crucial inputs in the discovery of the Higgs boson and in the search for new physics at the Large Hadron Collider (LHC). Many mid-energy facilities worldwide, such as a future electron-ion collider (EIC) \cite{
AbdulKhalek:2022hcn}
at Brookhaven, are trying to determine these PDFs and their three-dimensional generalizations. 

These parton distributions are defined as lightcone correlators. To compute them using lattice Quantum Chromodynamics (QCD) is highly non-trivial. Because to avoid the ``sign problem" in Monte-Carlo simulations, the lattice QCD is defined in the Euclidean space where the lightcone does not exist. Different approaches are developed to tackle this problem, including hadronic tensors \cite{Liu:1993cv}, axuiliary quarks \cite{Detmold:2005gg},  
 current-current correlators \cite{Braun:2007wv,Ma:2017pxb}, smeared operators \cite{Davoudi:2012ya}, large-momentum effective theory (LaMET)~\cite{Ji:2013dva,Ji:2014gla}, gradient flow \cite{Monahan:2015lha,Shindler:2023xpd},
 pseudo-PDFs \cite{Radyushkin:2017cyf},
operator product expansion (OPE) \cite{Chambers:2017dov}, and an alternative formulation of the deeply inelastic scattering structure functions \cite{Mueller:2019qqj}.

Among these approaches, LaMET
is the most explored. Since its first calculation in lattice QCD \cite{Lin:2014zya}, a lot of progress has been made \cite{Ji:2020ect}. LaMET relates
a PDF $f$ to its quasi-PDF $\tilde f$, which is the Fourier transform of an equal-time correlator, in the limit of infinite hadron momentum $P_z$ through a factorization theorem, which is an OPE in inverse powers of $P_z$:
\begin{align}\label{eq:lamet2}
    \tilde f(x,P_z) &=  \int_{-1}^1 {dy\over |y|} C\Big({x\over y}, {\mu \over y P_z}\Big)  f(y,\mu) \nonumber\\
    & + {\cal O}\Big( {\Lambda_{\mathrm{QCD}}^2\over (xP_z)^2}, {\Lambda_{\mathrm{QCD}}^2\over ((1-x)P_z)^2},\frac{M^2}{P_z^2}\Big)\,.
\end{align}
$x$ and $y$ are the momentum fractions of the parton in the hadron infinite momentum frame, with negative $x(y)$ corresponding to the PDF of the antiparton. $\mu$ is the renormalization scale in the modified minimum subtraction scheme ($\overline{\text{MS}}$), typically chosen to define PDFs. $\Lambda_{\mathrm{QCD}}\sim 200$ MeV is the scale when the strong coupling constant becomes of order one, and $M$ is the hadron mass. The power corrections $(M^2/P_z^2)^n$ can be corrected for all powers of $n$ \cite{Chen:2016utp}. 

In principle, the matching factor $C$, a Wilson coefficient, is computable in perturbative QCD. However, soft gluons cause the IR renormalon ambiguity \cite{Braun:2018brg}
of order $\Lambda_{\mathrm{QCD}}^2/ (xP_z)^2$ and $\Lambda_{\mathrm{QCD}}^2/ ((1-x)P_z)^2$ to be canceled by power corrections. This is because the large momentum expansion does not converge anymore when the momentum of the parton ($x P_z$) or the rest of the partons ($(1-x) P_z$) becomes soft. Therefore, LaMET cannot provide reliable PDF extractions near the endpoints of $x$.      

Interestingly, this restriction is no longer there if the PDFs are computed through lightcone correlators on a Minkowski lattice on a quantum computer. (For recent reviews on applying quantum computations to high energy physics, see \cite{Klco:2021lap,Bauer:2022hpo,Bauer:2023qgm,Beck:2023xhh}.) Then the PDF $f(x,1/a)$ with lattice spacing $a$ can be related to the PDF $f(x,\mu)$ in $\overline{\text{MS}}$ scheme as
\begin{align}\label{matching2}
    f(x,\frac{1}{a}) =  \int_{-1}^1 {dy\over |y|} \mathcal{C}\Big({x\over y}, {\mu a}\Big)  f(y,\mu) 
    + {\cal O}\Big( a \Lambda_{\mathrm{QCD}}\Big)\,,
\end{align}
where the $\ln{(\mu a)}$ term in $\mathcal{C}$ in one loop is just the familiar Altarelli-Parisi kernel. No renormalon ambiguity causes the breakdown of the power expansion in $a$. Because $\mathcal{C}$ compensates for the difference
$f(x,1/a) - f(x,\mu) \equiv \delta f(x,1/a,\mu)$, where the counterterm $\delta f$ imposes the lattice cutoff and dimensional regularization at the same time so that the combination $f(x,\mu)+\delta f(x,1/a,\mu)$ 
can have the dimensional regularization removed without any consequence. Since $\delta f$ has no soft modes, there is no IR renormalon ambiguity in $\mathcal{C}$ to be canceled by the power corrections. This conclusion is verified with direct computation using the typical dressed gluon propagator by a chain of vacuum-polarization diagrams. Technically, when the longitudinal parton momentum is taken to infinity as required in a lightcone PDF, the dressed gluon diagrams relevant for IR renormalon computation all vanish. 

In addition to having this intrinsic advantage over Euclidean space approaches, quantum computation has some other practical advantages as well. For example, lattice QCD computations of the proton non-valence (strange, charm, and gluon) PDFs suffer from the noisy  ``disconnected" diagrams. In a quantum computation, valence and non-valence PDFs are treated equally. In fact, in the calculation that we will present, the PDF receives both valence and non-valence contributions already in this first exploratory study. 

Higher-dimensional parton distributions, such as generalized parton distributions (GPDs), which require lightcone correlators of hadronic matrix elements of different initial and final states, and transverse momentum distributions (TMDs), which require matrix elements of single- and double-stapled Wilson lines with each staple containing two antiparallel lightcone Wilson lines connected by a transverse Wilson line at spatial infinity. The setup of the quantum computations of these quantities is straightforward. Also, similarly to the PDF calculations, non-valence contributions cost no more resources than the valence ones to compute. Since the longitudinal parton momentum is taken to infinity as in PDFs, the renormalon ambiguity, which restricts access to the whole distributions, is likely to be missing. These are promising advantages in quantum computation worth further exploring.

In this work, we use IBM quantum computers to calculate the PDF of the lightest bound state in the simplest quantum gauge theory---a vector positronium state in Quantum Electrodynamics with 1+1 dimensions(D) (QED$_2$). In QED$_2$, the electric potential between an electron-positron pair is linear. Therefore, the positronium is confined like a meson in QCD in 3+1 D. 

In QED$_2$, the theory is super-renormalizable. The UV cut-off of the lattice theory $1/a$ can be removed smoothly by taking $a \to 0$. The lightcone PDFs in the continuum and on the lattice are both finite and  
the matching formula of Eq.(\ref{matching2}) becomes
\begin{eqnarray}\label{eq:3}
    f(x,a) =    f(x) 
    + {\cal O}(a)\, .
\end{eqnarray}
Hence 
\begin{eqnarray}
f(x)=f(x,a\to 0) .
\end{eqnarray}


\section{Hamiltonian Formulation for QED$_2$}


We first discuss the formulation in the continuum, then on a spatial lattice.

\subsection{In the Continuum}
The QED$_2$ Lagrangian density is
\begin{equation}
\label{L}
\mathcal{L}=-\frac{1}{4}F_{\mu\nu}F^{\mu\nu}+\bar{\psi}(i \slashed{D}-m)\psi ,
\end{equation}
where the covariant derivative $D_{\mu}=\partial_{\mu}+i e A_{\mu}$. In 1+1 D, the mass dimensions of the operators and couplings are $[F^{\mu\nu}]=[\partial_{\mu}]=[e]=[m]=1$, $[\psi]=1/2$, and $[A_{\mu}]=0$. Naively $(\bar{\psi}\psi)^2$ and \Blue{$\bar{\psi}\sigma_{\mu\nu}F^{\mu\nu}\psi$} are also relevant operators of mass dimension two. However, 
if they are induced from terms of Eq.(\ref{L}), then there will be $e^2$ and $e$ in their prefactors, respectively, which make them irrelevant operators. (On the other hand, 
in a non-gauge theory, the $(\bar{\psi}\psi)^2$ term becomes a relevant operator.) 
The $\theta \epsilon_{\mu\nu}F^{\mu\nu}$ term is also a relevant operator which can be added to the Lagrangian density. This term breaks parity and charge conjugation separately. We will work with the system with $\theta=0$.

The conjugate momentum for the gauge field $A_{\mu}$ is
\begin{equation}
\Pi^{\mu}=\frac{\partial \mathcal{L}}{\partial(\partial_0 A_{\mu})}=F^{\mu 0}.
\end{equation} 
The fact that $\Pi^{0}=0$ implies $A^0$ is not dynamical. Integrating out $A^0$ from the path integral by completing the square of the Gaussian integral demands that the remaining integrand be evaluated with the constraint \cite{Weinberg1995}:
\begin{equation}
\frac{\partial \mathcal{L}}{\partial A_0}=J^0-\partial_{\mu} \Pi^{\mu} =0 ,
\end{equation}
which is Gauss's law --- a constraint holds even at the quantum level.

The conjugate momentum of the fermionic field is 
\begin{equation}
\pi=\frac{\partial \mathcal{L}}{\partial_0 \psi}=i\psi^{\dagger}.
\end{equation}
Imposing an open boundary condition in the $x$-direction such that $A^1$=0 (the axial gauge) can be achieved on all the gauge links, then the Hamiltonian density is
\begin{eqnarray}
\label{H2}
\mathcal{H}&=&\pi \partial_0 \psi-\mathcal{L} \nonumber \\ 
&=&\frac{1}{2}E^2-i \bar{\psi}\gamma^1 \partial_1 \psi+m \bar{\psi}\psi,
\end{eqnarray}
where we have dropped the $A^0 (J^0-\partial_x E)$ term by applying the Gauss's law.

We are interested in lightcone correlators defined as the matrix element of the operator 
\begin{eqnarray}
\label{Oz}
\mathcal{O}(z_1,z_2)&=&  \bar{\psi}(z_1 n) n\cdot \gamma W_n(z_1 n,z_2 n)\psi(z_2 n) ,
\end{eqnarray}
where $n^{\mu}=(1,-1)$ is a lightlike vector and the gamma matrices $\gamma^0 n\cdot \gamma=\gamma^0(\gamma^0+\gamma^1)=\begin{pmatrix}
1 &-1 \\
-1 &1
\end{pmatrix}$. The Wilson line
\begin{equation} 
W_n(z_1 n,z_2 n)=
e^{-ie\int^{z_1}_{z_2} n \cdot A(z)d z}
\end{equation}
makes the operator of Eq.(\ref{Oz}) gauge invariant.

One can show that
\begin{eqnarray}
\label{heritian}
\mathcal{O}^{\dagger}(z_1,z_2)
&=& \mathcal{O}(z_2,z_1) ,
\end{eqnarray}
and under charge conjugation operation $\mathcal{C}$, 
\begin{eqnarray}
\label{C}
   \mathcal{O}(z_1,z_2) \overset{\mathcal{C}}{\to} -\mathcal{O}(z_2,z_1).
\end{eqnarray}

For a state $|P \rangle$ with momentum $P$, the matrix element of a state with momentum $P$
\begin{eqnarray}
\label{Dz}
   D(z_1-z_2) 
    &=& \frac{1}{2 n\cdot P}\langle P|\mathcal{O}(z_1,z_2)|P \rangle 
\end{eqnarray}
is invariant under spacetime translation and depends on $z\equiv z_1-z_2$ only. 
However, translational symmetry in the $x$-direction is broken by the open boundary condition. We need the lattice size to be much bigger than the size of the (zero-momentum) positronium to make this finite volume effect under control. 

Eq.(\ref{heritian}) implies the real(imaginary) part of $D(z)$ is even(odd) in $z$. We will focus on the case that the positronium state $|P \rangle$ is also an eigenstate of $\mathcal{C}$, such that
the real part of $D(z)$ vanishes according to Eq.(\ref{C}). Therefore, the PDF of the positronium
\begin{eqnarray}
\label{f(x)}
f(x)&=& \int_{-\infty}^{\infty} \frac{dz n\cdot P}{2 \pi} e^{-i z n\cdot P x} D(z) 
\end{eqnarray}
is real but odd in $x$. $f(x)$ is the probability density of finding a parton of the momentum fraction $x$ of the positronium in the infinite momentum frame of the positronium. However, $D(z)$ in Eq.(\ref{Dz}) is a Lorentz scalar which can be computed in any Lorentz frame. We will perform the calculation in the rest frame of the positronium.  

In Eq.(\ref{f(x)}), the support of $x$ is $[-1,1]$, 
with the negative $x$ part corresponding to the anti-fermion ($\bar{q}$) PDF with $f_{\bar{q}}(|x|) = -f_q(-|x|)$.
Therefore, the first moment of the PDF yields the net number of fermions $q$ in the positronium
\begin{eqnarray}
   \int_{-1}^{1}dx f(x) =\int_{0}^{1} dx[f_q(x)-f_{\bar{q}}(x)]= 0 .
\end{eqnarray}
The second moment of the PDF yields the momentum fraction of the positronium carried by $q$ and $\bar q$. The gauge field does not carry momentum because its contribution to the momentum density (or the $(01)$ component of the energy momentum tensor) is proportional to $F^{0\mu}(x)F^{\ 1}_{\mu}(x)=0$. Therefore, all the positronium momentum is carried by the fermion: 
\begin{eqnarray}
\label{<x>}
   \int_{-1}^{1}dx x f(x) =\int_{0}^{1} dx x[f_q(x)+f_{\bar{q}}(x)]= 1.
\end{eqnarray}

 
Unlike the first two moments, which are conserved quantities, one can also look at the number of the fermion anti-fermion pairs in the positronium:
\begin{eqnarray}
   \int_{0}^{1}dx f(x) .
\end{eqnarray}
This is not a conserved quantity. It depends on $a$ in our case. 


The computation of the lighcone correlator of Eq.(\ref{Dz}) requires treating the lightlike Wilson line as carrying the same charge as the fermion when applying Gauss's law. This can be shown by applying the auxiliary field approach, which recasts the Wilson line as a ``heavy particle" propagator in coordinate space \cite{Chen:2016fxx,Ji:2017oey}. To show this, we replace the operator in Eq.(\ref{Oz}) by 
\begin{equation}\label{Eq:repl}
             O'(x, y) = \overline{\psi}(x)n\cdot\gamma Q_n(x) \overline{Q}_n(y) \psi(y) ,
\end{equation}
where $Q_n$ carries the same charge as $\psi$ such that $\overline{\psi}(x)n\cdot\gamma Q_n(x)$ and $\overline{Q}_n(y) \psi(y)$ are both gauge invariant. In addition, one term in the Lagrangian is added which resembles a ``heavy fermion" field with its ``2-velocity" along the lightcone direction $n$:
\begin{eqnarray}
\label{Eq:lag}
\label{LQ}
{\cal L'}={\cal L}+\overline Q_n(x)i n\cdot D Q_n(x).
\end{eqnarray}

Integrating out the heavy fermion field, one has 
\begin{eqnarray}
    \int{\cal D}\overline Q_n{\cal D}Q_n\, Q_n(x)\overline Q_n(y) e^{i\int d^2x {\cal L'}} 
    =S_n(x, y) e^{i\int d^2x {\cal L} },
\end{eqnarray}
where the Green's function satisfies 
\begin{eqnarray}
n\cdot D\, S_n(z_1 n, z_1 n)=\delta^{(2)}(x-y) 
\end{eqnarray}
has the solution
\begin{align}
&S_n(x,y)=\theta(x^+-y^+)\delta(x^--y^-)W_n(x^+,y^+) 
\end{align}
with $x^{\pm}=(x^0\pm x^3)$. Therefore, the Wilson line can be viewed as a heavy fermion propagator carrying the same charge of $\psi$. This charge contributes to Gauss's law by contributing a new term to the current  
\begin{eqnarray}
\delta J^{\mu}=-e \overline Q_n(x) n^{\mu} Q_n(x) 
\end{eqnarray}
through Eq.(\ref{LQ}).
 
\subsection{On a Lattice}

For the Hamiltonian density of Eq.(\ref{H2}) in axial gauge with an open boundary condition in the $x$-direction, we use staggered fermions and the Jordan-Wigner transformation to map the Hamiltonian to its qubit form \cite{Sala:2018dui,Savage2025}:
\begin{eqnarray}
\label{H}
H 
&=& 
\frac{1}{2a} \sum_{j=-L+1}^{L-1} \Big[ \sigma^+_j \sigma^-_{j+1} + \text{h.c.} \Big]+\frac{m}{2} \sum_{j=-L+1}^{L} \Big[(-1)^j Z_j + I \Big]\nonumber \\
&+& \frac{e^2a}{2} \sum_{j=-L+1}^{L-1} \Big[\sum^j_{k=-L+1}  Q_k \Big]^2 ,
\end{eqnarray}
where Gauss's law 
\begin{equation}
\label{Gauss}
E(x_k)-E(x_{k-1})- e Q_k =0
\end{equation}
has been used with the boundary value $E(x_{-L})=0$ that is consistent with $\theta=0$, and where
\begin{eqnarray}
Q_k = -\frac{1}{2} \Big[Z_k + (-1)^k I \Big].
\end{eqnarray}
One should work in the limit of $m,e \ll 1/a$ with $1/a$ the UV cut-off scale. 
We will set the lattice spacing between two staggered sites $a=1$ from now on.

The lightcone operator of Eq.(\ref{Oz}) with $(z_1,z_2)=(z,0)$ and $z=2k$ can be written as \cite{Li:2021kcs}
\begin{eqnarray}
\label{OO1}
\mathcal{O}(z,0)
&=&e^{i2 k H}\Bigl\{\mathcal{Z}^{*}_{-2k-1}\sigma^+_{-2k} e^{-i2 k H} W_{1,n} \mathcal{Z}_{-1}\sigma_{0}^-\nonumber\\
&+&\mathcal{Z}^*_{-2k}\sigma^+_{-2k+1} e^{-i2 k H} W_{2,n}\mathcal{Z}_{0}\sigma_{1}^-\nonumber\\
&-&\mathcal{Z}^*_{-2k-1}\sigma^+_{-2k} e^{-i2 k H} W_{3,n}\mathcal{Z}_{0}\sigma_{1}^-\nonumber\\
&-&\mathcal{Z}^*_{-2k}\sigma^+_{-2k+1}e^{-i2 k H} W_{4,n}\mathcal{Z}_{-1}\sigma_{0}^-\Bigr\},
\end{eqnarray}
where 
\begin{eqnarray}
\mathcal{Z}_k=(-iZ_{-L+1})\cdots (-iZ_{k-1})(-iZ_{k})
\end{eqnarray}
from Jordan-Wigner transformation. The $W_{1-4,n}$ Wilson lines can end at even or odd staggered sites and make $\pm 90$ degree zigzag turns instead of going straight in 45 degrees in the spacetime diagram. In the axial gauge, $W_{1-4,n}$ becomes identical. As explained above, the Wilson line carries the same charge as the fermion. (This is also obtained in \cite{Schneider:2024yub,Banuls:2025wiq} using the temporal gauge ($A^0=0$).) In the axial gauge, the Wilson line is the multiplication of gauge links in the time direction with 
\begin{equation}
\label{A0}
A^0(x_k)=-\sum_{m=-L+1}^k Q_m(k-m)+ c k + d ,
\end{equation}
which yields the expected relation $A^0(x_k)-A^0(x_{k-1})=-E(x_{k-1})$. When $\theta=0$, the background E field $-c=0$. The constant $d$ can be absorbed by the fermionic fields in the lightcone operator. This construction can be readily generalized to higher dimensional non-abelian theories without having to take a derivative on each time slice as suggested in Ref. \cite{Lamm:2019uyc}. For a Wilson line in the $t$-$z$ direction, one can take the axial gauge $A^3_b=0$ under an open boundary condition, with $b$ an internal space index. Then $A_b^0$ can be solved from $E_b^3\equiv F_b^{03}=-\partial^3 A_b^0$. 

On a lattice, the continuum definition of Eq.(\ref{Dz}) needs to be changed to
\begin{eqnarray}
   D(z_1-z_2) 
    &=& L\langle P|\mathcal{O}(z_1,z_2)|P \rangle ,
\end{eqnarray}
where $L$ is the number of physical spatial sites (which is half of the staggered sites in 1+1 D) such that Eq.(\ref{f(x)}) remains unchanged. In a discrete theory, the state $|P \rangle$ is normalized to unity and the operator $\mathcal{O}(0,0)$ is the local current density whose matrix element is proportional to $1/L$. 
In the continuum limit, the normalization of $\langle h|\mathcal{O}(z,0)|h \rangle$ has an extra factor of $2 n\cdot P L a$.

The lattice artifact from using a Zigzag Wilson line instead of a straight one 
can be investigated by a lattice perturbation theory calculation. The loop integrals are all finite even if $a\to 0$ is taken before the integrations are performed---a consequence of QED$_2$ being super-renormalizable. This leads to Eq.(\ref{eq:3}).

\section{Quantum Computations}
\label{QC}

We perform the calculation with IBM Eagle R3 
quantum computers using superconducting qubits. The main source of error is from two-qubit gate errors, while the one-qubit gate errors are much smaller in comparison. At the moment, IBM computers cannot handle quantum circuits with more than 5,000 two-qubit gate depth. 
We found it critical to reduce the two-qubit gate depth as much as possible to get sensible signals.  

We use 10 qubits to describe a system of 10 staggered fermion sites (corresponding to 5 physical spatial points) plus one ancillary qubit put at the center by choice to perform matrix element computations following the prescription of \cite{PhysRevLett.113.020505,Li:2021kcs}. 
Our original program, which has a gate depth close to 5,000, does not yield a sensible result. We have taken the following steps to reduce the gate count by one order of magnitude so that sensible results finally emerge after error mitigation.

(1) State preparation: We use wave packets instead of exact diagonalization eigenstates to reduce the gate count. We have tuned the wave packet parameters of \cite{Farrell:2023fgd,Farrell:2024fit} to increase the overlap with the eigenstate from exact diagonalization. However, there is a trade-off---a larger overlap requires a more elaborate construction of the wave packet, and hence a larger gate depth. We use 65.1\% overlap, which reduces the two-qubit gate depth to 112 from 1751 with exact diagonalization.


 (2) Operator choice: The discussion of 
 Eq.(\ref{Dz}) shows all the
 $\mathcal{O}(z+\Delta z,\Delta z)$ operators yield the same matrix element $D(z)$ provided $|P \rangle$ is an energy momentum eigenstate. However, the wave packet that we employ is only an approximated zero momentum eigenstate with its center of mass at the origin. Hence, naively, choosing $\Delta z=-z/2$
 to use the $\mathcal{O}(z/2,-z/2)$ operator should better preserve the symmetry of Eq.(\ref{heritian}). However, non-zero $\Delta z$ will increase the required two-qubit gates through time translation operation such that $\mathcal{O}(z,0)$ defined in Eq.(\ref{OO1}) costs $\sim 66\%$ less two-qubit gates than $\mathcal{O}(z/2,-z/2)$. 
Therefore, we use the operator $[\mathcal{O}(z,0)+\mathcal{O}(0,z)^{\dagger}]/2$ to reduce the gate count and preserve Eq.(\ref{heritian}). This yields a symmetrized matrix element 
\begin{eqnarray}
\label{tildeD}
    D(z) \to\frac{D(z)+D^*(-z)}{2}
\end{eqnarray}
such that the real(imaginary) part of $D(z)$ is even(odd) in $z$. 

(3) Wilson Line and time evolution operator: We use zigzag gauge links in the time and spatial directions to approximate the Wilson line along the lightcone. We change the length of each gauge link and time evolution step from $a$ to $2a$ with 
the second-order Trotterization and 
Trotter step = 1 \cite{Farrell:2024fit}. This reduces the two-qubit gate depth by about half with the change of $D(z)$ well within our error bars ($|\delta D(z)|\sim 0.1$ for $|z|=2,4$ from a classical simulator). However, the error of using zigzag Wilson lines is not systematically studied in this work.  

(4) Error mitigation: We use the built-in functions of dynamical decoupling (DD) \cite{Viola_1998,Souza_2012,Ezzell:2022uat} and Pauli twirling (PT) \cite{Wallman:2015uzh} in qiskit and use randomization = 100 for PT (which means sampling 100 possible twirling combinations to approximate all possible combinations).
The Operator Decoherence Renormalization (ODR) \cite{ARahman:2022tkr} is computed for the positronium matrix element of the operator ($z=2k$)
\begin{eqnarray}
\label{OO2}
\mathcal{O}'(z)
&=&e^{i k H}e^{-i k H}\nonumber\\
&\times& \Bigl\{\sigma^+_{0} e^{i k H} e^{-i k H} W_{1,n}^{-\frac{1}{2}} W_{1,n}^{\frac{1}{2}} \sigma_{0}^-+\cdots\Bigr\} ,
\end{eqnarray}
which is similar to $\mathcal{O}(z,0)$ of Eq.(\ref{OO1}) 
but with the Wilson line going forward in time along the lightcone direction then coming back to where it started. The matrix elements of $\mathcal{O}(z,0)$ and $\mathcal{O}'(z)$ have similar two-qubit gate counts---the difference is less than 8\% for $z=4$ and both have DD and PT(randomization=100) implemented. The ODR matrix element $L\langle P|{O}'(z)|P \rangle$ would be one if there were no errors. The deviation from one is used to calibrate the normalization of $D(z)$.

\begin{figure}[t!]
\includegraphics[width=\columnwidth]{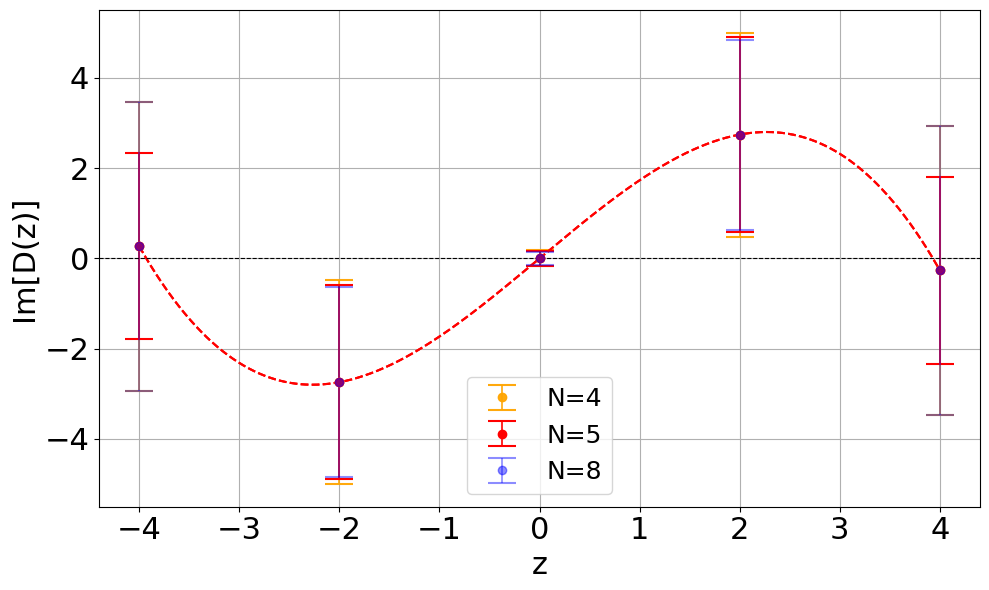}
        \\
\includegraphics[width=\columnwidth]{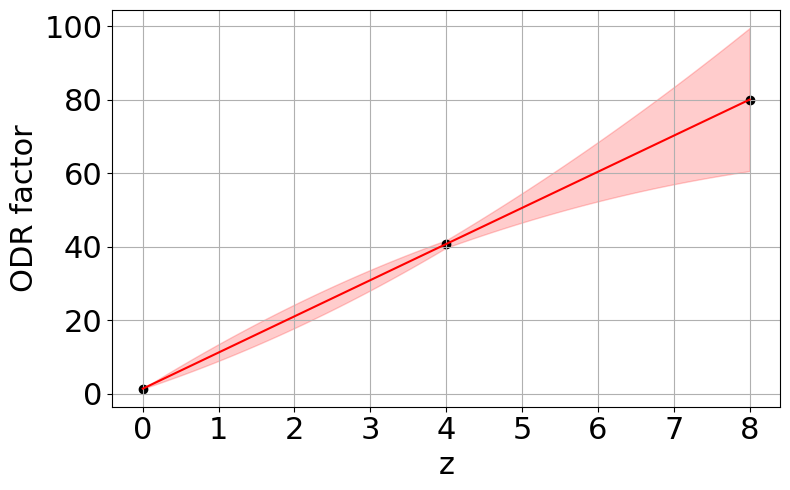}
\caption{(Upper) The imaginary part of $D(z)$ for lattice size $2La=10a,(e a,m a)=(0.3,0.5)$ with DD, PT and ODR error mitigation. Local phase rotations (see step (5) of Sec. \ref{QC}) and symmetrization (see Eq.(\ref{tildeD})) are applied such that $D(z)$ becomes purely imaginary and odd in $z$. The standard error is insensitive to the batch number $N$ that the 4,000 shot results are distributed into. The results for $N=4,5,8$ are shown.  
(Lower) The ODR normalization enhancement factor used in the upper plot is shown as a function of the length of the Wilson line $z$.}
    \label{fig1}
\end{figure}

(5) Violation of charge conjugation $\mathcal{C}$: Lattice discretization gives rise to $\mathcal{C}$ violation because fermions and anti-fermions are put on different staggered sites. As a consequence, the real part of $D(z)$ defined in Eq.(\ref{tildeD}) does not vanish. Although we see Re[$D(z)$] indeed becomes smaller with smaller lattice spacing on a classical simulator, we do not have the resources yet to perform the continuum limit extrapolation on a quantum computer. Instead, we parametrize the error as transforming $D(z)$ to $\rho(z) e^{i\phi(z)}D(z)$, with real $\rho$ and $\phi$. $\phi(z)$ comes from the $\mathcal{C}$ violating effects while $\rho(z)$ comes from both $\mathcal{C}$ violating and conserving effects. While this parameterization is general, we will assume that the overall error factor is the product of individual two-qubit gate error factors: 
$\rho e^{i\phi}=\Pi_j \rho_j e^{i\phi_j}$.
Therefore, if our ODR operator of Eq.(\ref{OO2}) have the same number of two-qubit gates as the operator for $D(z)$ in Eq.(\ref{OO1}), then the ODR matrix element will be able to remove the factor $\rho(z)$ but not the phase $\phi(z)$ (the error phase accumulated from evolution forward in time is different from evolution forward then backward in time). We remove the phase in $D(z)$ by performing a $z$-dependent local phase rotation to make $D(z)$ purely imaginary. We then remove the phase in ODR by a different local phase rotation to make it real and positive. So the final ODR factor presented in Fig.\ref{fig1} would be $1/\rho(z)$ if the two-qubit gate counts between operators of Eqs.(\ref{OO1}) and (\ref{OO2}) are the same. In reality, there is $\sim 8\%$ difference in the gate counts for $|z|=4$, partly due to the Pauli $\mathcal{Z}_k$ factors that form controlled $Z$ gates with the ancillary qubit. We find the gate counts scale linearly with $z$ and so does the ODR factor (see Fig.\ref{fig1}). Therefore, the gate count difference implies that the ODR factor should receive a $\sim 8(4)\%$ correction at $|z|=4(2)$, which is much smaller than errors of $D(z)$ and can be neglected. 

In our study, step (1) above reduces the two-qubit gate depth from about 5,000 to $3,400$, step (2) reduces it to about $1,100$, and then step (3) reduces it to approximately $500$. The DD and PT error mitigations are one-qubit gate operations, so they do not change this counting. We have not explored the possibility of further reducing gate counts while maintaining the same precision in this exploratory study. 


Now we present our numerical results. In the upper plot of Fig.\ref{fig1}, the imaginary part of $D(z)$ for lattice size $2La=10 a$, $(e a,m a)=(0.3,0.5)$ is shown. Local phase rotations (step (5) of Sec. \ref{QC}) and symmetrization (Eq.(\ref{tildeD})) are applied such that $D(z)$ becomes purely imaginary and odd in $z$. Error mitigation is applied using DD, PT(randomization=100), and ODR (step (4) of Sec. \ref{QC}). The result is based on 4,000 shot calculations of $D(z)$ with qiskit built-in functions of DD and PT together with 16,000 shot calculations of the ODR factor, both run on the \texttt{ibm\_sherbrooke} quantum computer. The bitstring result is divided into $N=4,5,8$ batches to compute the standard error. If the underlying error distribution is Gaussian, the standard errors would be $N$ independent. This seems to be the case except for the points at the boundaries where the errors of $N=4$ and 8 are about 50\% larger than those of $N=5$.     

In the lower plot of Fig.\ref{fig1}, the ODR normalization enhancement factor used in the upper plot is shown as a function of the length of the Wilson line, $z$. The $z=2$ result is interpolated with quadratic fits.

\begin{figure}[t!]
\includegraphics[width=\columnwidth]{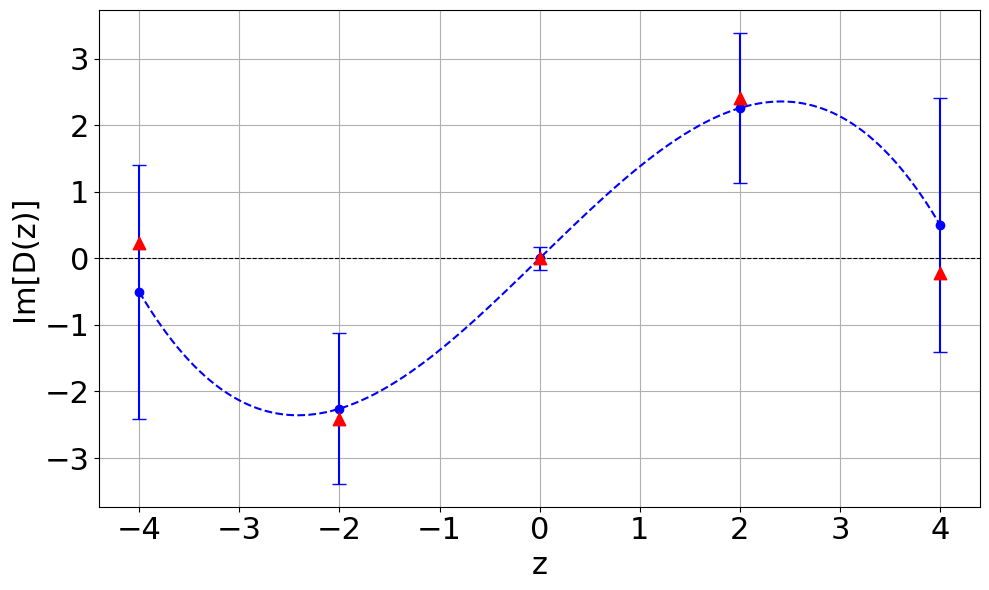} \includegraphics[width=\columnwidth]{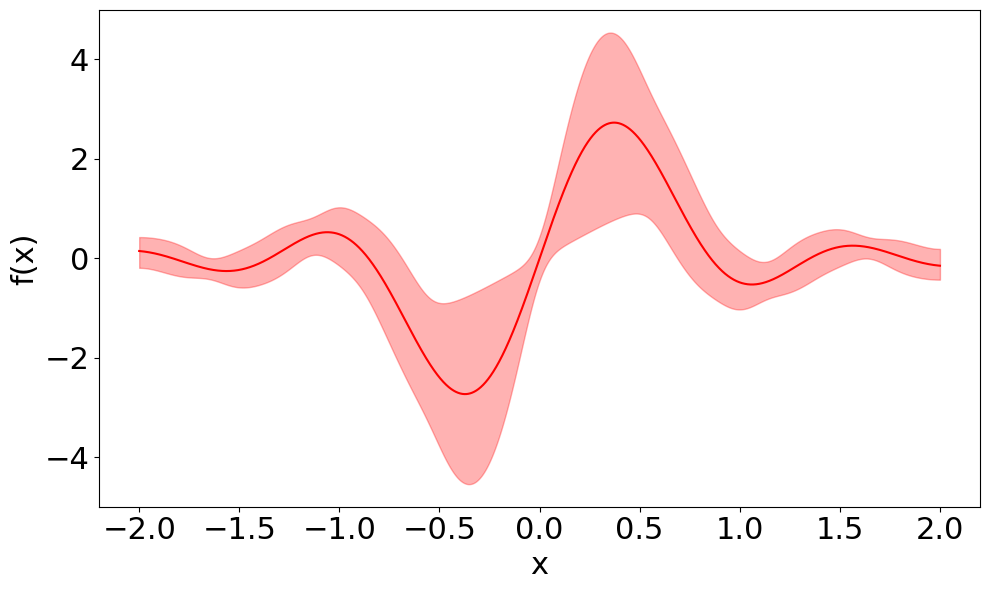}
\caption{(Upper) Combined result of the imaginary part of $D(z)$ with DD+PT+ODR from \texttt{ibm\_sherbrooke} and \texttt{ibm\_kyiv} with 4,000 shots each. The triangles are results from a classical simulator. 
(Lower) The $f(x)$ result from the Fourier transform of the $D(z)$ result above.
    }
\label{fig2}
\end{figure}

In the upper plot of Fig.\ref{fig2}, the 4,000 shot $N=4$ result of $D(z)$ 
in Fig.\ref{fig1} is 
combined with another 4,000 shot result from \texttt{ibm\_kyiv} using the Cochran(1977) weighted standard error formula of Ref.\cite{GATZ19951185}. This formula yields results consistent with the bootstrap method. The combined result is consistent with Fig.\ref{fig1}, but with standard errors reduced by about one-half. This is because the standard error, SE, is related to the standard deviation, $\sigma$, through $\text{SE}=\sigma/\sqrt{N_{\text{shots}}}$ with $N_{\text{shots}}$ being the number of times (shots) the quantum computation is carried out.
Since $\sigma\propto 1/\sqrt{N_{\text{shots}}}$, then SE $\propto 1/N_{\text{shots}}$. 
This result is consistent with the classical simulator result marked by triangles, with their central values almost coinciding for points away from the boundary. This suggests that our error mitigation is working, although the errors are still large.

In the lower plot of Fig.\ref{fig2}, we take the Fourier transform of the combined $D(z)$ result to obtain the PDF $f(x)$ using Eq.(\ref{f(x)}) and the positronium mass
$M a=1.617\pm 0.002$ 
from the difference of the Hamiltonian expectation values between the first excited state and the ground state (vacuum) using a classical simulator.
The distribution outside the physical region $x=[-1,1]$ is consistent with zero. The total momentum carried by the partons is
\begin{eqnarray}
   \int_{-1}^{1}dx x f(x) &=& 1.04^{+0.66}_{-0.76} ,
\end{eqnarray}
where the errors are asymmetric because the positivity of $xf(x)$ is imposed in the error computation. It is encouraging that the central value agrees with the sum rule of Eq.(\ref{<x>}) at the 4\% level, although the error is still large.  

Our result also yields the number of fermion anti-fermion pairs in the positronium:
\begin{eqnarray}
\label{pair}
   \int_{0}^{1}dx f(x) &=&1.37^{+0.89}_{-1.02} ,
\end{eqnarray}
where the positivity of $f(x)$ for $x=[0,1]$ is imposed in the error calculation.
As mentioned above, this integral is not a conserved quantity. Its value depends on $e/m$ and the renormalization scheme and scale. In our QED$_2$ calculation, $f(x)$ depends on $1/a$, although the theory is super-renormalizable. In the heavy fermion and continuum limit ($e/m \to 0$ and  $a \to 0$), one expects that there is just one fermion pair in the lightest positronium since the potential energy is much smaller than $2m$, pair production is highly suppressed. 
For finite $e/m$, one expects $\int_{0}^{1}dx f(x) \ge 1$. At finite $a$ and finite volume, this bound might be modified \cite{Banuls:2025wiq}. 
Our central value is consistent with this bound, although the error is large, while
the results of the 1+1 D Nambu-Jona-Lasinio model are below this bound \cite{Li:2021kcs,Kang:2025xpz}. 

In our calculation, all the partons, valence or not, are accounted for since they all show up in the positronium state. There is no
difference
in the resource requirement to compute the valence or nonvalence PDFs, which is an interesting distinction from lattice QCD approaches.

In this work, the errors presented are all statistical. As discussed above, the standard errors scale as $1/N_{\text{shots}}$, which can be reduced with more resources. For example, if $N_{\text{shots}}=160,000$ (20 times more than this work) is available, then the momentum sum rule statistical error can be reduced to 4\%. 

Systematic errors, such as the discretization errors and the finite volume effect, require more advanced quantum computers to reduce them. Numerically, we find the number of two-qubit gate depths scales as $L^3$, with $L$ the physical size, in the $D(z)$ computation. This factor comes from having $\mathcal{O}(L)$ time evolution steps, each step has $\mathcal{O}(L^2)$ two-qubit couplings in the Hamiltonian. For our $L=5$ calculation, the two-qubit gate depths for $z=\{-4,-2,0,2,4\}$ is
$\{529, 367, 112, 319, 615\}$, respectively. With the limitation of 5,000 gate depths on current machines, one is limited to $L<10$. However, with the Starling machines scheduled to be available in 2029 according to the IBM Quantum Roadmap,  $L\sim 100$ becomes possible. The Starling's 100M two-qubit gate capacity
can support up to $L=273$ calculations. The limiting factor comes from the qubit number of 200, which sets a limit of $L=99$ in the calculation. With the next-generation Blue Jay machines with 1B two-qubit gates and 2,000 qubits available in 2033+, $L=588$ is possible. It would be interesting to explore what can be achieved for QCD in 3+1 dimensions with these machines. 

\section{Conclusions}

We have carried out the first quantum computation of PDFs with a real quantum device. 
We calculate the PDF of a positronium in QED in 1+1 dimensional spacetime with IBM quantum computers. The calculation uses 11 qubits for a lattice system of 5 spatial sites (or 10 staggered sites) and one ancillary qubit. We have applied several error mitigations including the by now standard DD, PT and ODR procedures, and some steps specially developed in this work to preserve the symmetries of the problem. The most critical and challenging part, is to reduce the number of two-qubit gates from close to 5,000 (the maximum number executable in Eagle R3 machines) in the initial stage to around 500, such that sensible results start to emerge. 

The resulting lightcone correlators have excellent agreement with the classical simulator result in central values except at the spatial boundary, although the error is still large. 
The parton momentum is $1.04^{+0.66}_{-0.76}$ times of the positronium momentum, with the central value very close to the expected result of unity, although the error is still large. The number of electron-positron pairs in our positronium state is $1.37^{+0.89}_{-1.02}$. This quantity depends on the resolution scale, or the lattice spacing, in the computation. But the result suggests that not only valence partons have contributed to the computed PDF. The errors listed are only statistical. They scale inversely with the number of shots and can be further reduced with more resources. 

With the planned Starling (Blue Jay) machines, calculations with 20(117) times more sites than our system become possible. The tensor network result of Ref. \cite{Banuls:2025wiq}  suggests that our finite volume effect can be largely(completely) removed if we double(triple) the volume, and if the lattice artifact is incentive to $e/m$, then it can be removed if we reduce $a$ by half. So, the remaining systematics are expected to be removed if we increase the site number by a factor of 6.
However, for 1+1 D problems, classical approaches, such as the tensor network \cite{
Schneider:2024yub,
Kang:2025xpz, Banuls:2025wiq} and exact diagonalization \cite{Grieninger:2024cdl}, could already be very powerful. Hence, a quantum advantage is expected to come from problems of higher dimensions. 

We have also explored what advantage a quantum computer can have over the most powerful classical lattice QCD approach, LaMET. The feature of being free from renormalon ambiguity, such that there is no restriction on the accessible momentum fraction $x$, is probably the most significant. Also, accessing non-valence PDFs without the difficulty of disconnected diagrams could be practically important. These advantages are likely to be carried over to higher-dimensional parton distribution studies such as GPDs and TMDs. 

Finally, a PDF calculation with 3+1 dimensional QCD near $x=0$ or $x=1$ will be a clear demonstration of the quantum advantage in a problem with great scientific impact \cite{Lin:2017snn}.


\ \ \

\section*{Acknowledgement}
JWC thanks Yi-Xian Chen for the renormalon discussion
and members of the InQubator for Quantum Simulation (IQuS) for stimulating discussions during his sabbatical leave. GM thanks Manuel Schneider for the discussions on 
Refs. \cite{Schneider:2024yub,Banuls:2025wiq}.
We thank Martin Savage for his early participation in this work and his very useful lecture notes on
``Quantum Information and Simulation for Scientific Applications". We thank the IBM Quantum Hub at NTU for providing computational resources and access to perform calculations on IBM quantum systems. This work is supported in part by the National Science and Technology Council of Taiwan under
Grant No. 113-2112-M-002-012. 

\bibliography{QCPDF.bbl}

\end{document}